\documentclass[11pt,letter,onecolumn]{article}
\usepackage[utf8]{inputenc}
\usepackage[english]{babel}
\usepackage{amsmath}
\usepackage{amsfonts}
\usepackage{amssymb}
\usepackage{makeidx}
\usepackage{graphicx}
\usepackage{lmodern}
\usepackage{fourier}
\usepackage[left=2cm,right=2cm,top=2cm,bottom=2cm]{geometry}
\author{Johnny Alejandro Mora Grimaldo\\
Gabriel T\'ellez \\
Departamento de F\'{\i}sica, Universidad de los Andes, Bogotá, Colombia}
\title{Relations Among Two Methods for Computing the Partition
  Function of the Two-Dimensional One-Component Plasma}
\date{}
\begin{document}
\maketitle
\begin{abstract}
The two-dimensional one-component plasma ---2dOCP--- is a system
composed by $n$ mobile particles with charge $q$ over a neutralizing
background in a two-dimensional surface. The Boltzmann factor of this
system, at temperature $T$, takes the form of a Vandermonde
determinant to the power $\Gamma = q^2/(2\pi\varepsilon k_BT)$, where
$\Gamma$ is the coupling constant of this Coulomb system.  The
partition function of the model has been computed exactly for the even
values of the coupling constant $\Gamma$, and a finite number of
particles $n$, by two means: 1) by recognizing that the Boltzmann
factor is the square of a Jack polynomial and expanding it in an
appropriate monomial base, and 2) by mapping the system onto a
1-dimensional chain of interacting fermions. In this work the
connection among the two methods is derived, and some properties of
the expansion coefficients for the power of the Vandermonde
determinant are explored.  
\end{abstract}

\section{Introduction}

The Two-Dimensional One-Component Plasma ---2dOCP--- is a model in
statistical physics in which $n$ mobile particles with charge $q$ are
distributed in a two-dimensional space, over a background of
neutralizing charge. The statistical treatment of this model begins
with the solution of the two dimensional Poisson's equation, from
which the Hamiltonian of the system is constructed. In a
two-dimensional universe, the electrostatic interaction among two
point-like charges $q$, located at positions $\textbf{r}_i$ and $\textbf{r}_j$ is given by the potential: 
\begin{equation}
U(\textbf{r}_i,\textbf{r}_j)=-\frac{q^2}{2\pi\varepsilon}\ln\frac{\left|\textbf{r}_i-\textbf{r}_j\right|}{L}.
\end{equation}

Working on \textit{cgs} units, we have $2\pi\varepsilon = 1$. $L$ is
an irrelevant arbitrary length that will be set to 1 in the
following. If the charges are distributed over a uniformly charged
disk of radius $R$, with background charge density $-nq/(\pi R^2)$, the potential energy of the system ---built up
from the background-background interaction, the background-particle
interaction and the particle-particle interaction---
is~\cite{er2d,oc2d}
\begin{equation}
U =  \frac{nq^2}{2R^2} \sum_{i=1}^{n}{r_i^2}-q^2\sum_{i<j} \ln \left| \mathbf{r}_i - \mathbf{r}_j \right| + U_0^D,
\label{ecu:potential_energy_total}
\end{equation}
where $U_0^D$ accounts for constant ---position-independent--- terms. The charges may as well be located at the surface of a uniformly charged sphere of radius $R$. Making use of the Cayley-Klein parameters of the sphere
\begin{eqnarray}
&u_k=e^{i\phi_k/2}\cos(\vartheta_k/2)&\\
&v_k=-ie^{-i\phi_k/2}\sin(\vartheta_k/2),&
\end{eqnarray}
where $(\vartheta_k,\phi_k)$ are the usual spherical coordinates of the
$k$-th particle,
we have~\cite{ers}
\begin{equation}
U = -q^2\sum_{i < j}\ln\left(2R\left|u_iv_j-v_iu_j\right|\right) + U_0^S.
\end{equation}
Here, again, $U_0^S$ stands for the constant terms in the interaction.

Now me may proceed with the general expression for the partition
function of the system, specifically, with the configurational integral concerning the electrostatic interaction. For the plane, it is found that \cite{afr}:
\begin{equation}
Z_U = \frac{K}{n!}\int d\mathbf{r}_{1} \cdots \int d\mathbf{r}_n \, \left|\det(V)\right|^{\Gamma} \prod_{i=1}^{n} e^{-\Gamma n r_i^2/2R^2},
\end{equation}
where $\left[V\right]_{i,j}=z_{i}^{j-1}$ is a Vandermonde matrix in
the variables $z_i = r_ie^{i\theta_i}$, and $(r_i,\theta_i)$ being the
polar coordinates of the \textit{i}-th particle in the plasma, and $K$
accounts for the exponential term $\exp\left({-U_0^D/(k_BT)}\right)$,
where $k_B$ is the Boltzmann constant.

The region of integration is dependent on the model being analyzed. In the soft disk approach, the particles are free to move over the whole $\mathbb{R}^2$ region, whereas in the hard disk model the particles are confined inside a disk of radius $R$. The integration limits are adjusted to meet this requirements.

For the sphere, is useful to work with the stereographic projection
from its south pole to a plane tangent to its north pole. Describing
the position of each point on the sphere by the polar coordinates
$(r_i, \theta_i)$ of its stereographic projection on the plane, the calculations lead to~\cite{forresterbook}:
\begin{equation}
Z_U=\frac{K}{n!}\int d\textbf{r}_{1} \cdots \int d\textbf{r}_n \, \prod_{i=1}^{n} \frac{1}{\left[1+r_i^2/(4R^2)\right]^{2+(n-1)\Gamma/2}} \prod_{i < j}\left|z_i-z_j\right|^{\Gamma},
\end{equation}
where the integration is to be done over the whole plane, and $K = K'(2/R^2)^n = (2/R^2)^n\exp\left({-U_0^S/k_BT}\right)$. Expressing again the product of differences as a Vandermonde determinant we get:
\begin{equation}
Z_U=\frac{K}{n!}\int d\textbf{r}_{1} \cdots \int d\textbf{r}_n \,  \prod_{i < j}\left|\det(V)\right|^{\Gamma}\prod_{i=1}^{n} \frac{1}{\left[1+r_i^2/(4R)^2\right]^{2+(n-1)\Gamma/2}}.
\end{equation}
Let $g(r)$ be a function which depends only on the distance $r$ to the center. For the plane
\begin{equation}
g(r)=e^{-\Gamma n r^2/2R^2}
\end{equation}
whereas for the sphere
\begin{equation}
g(r)=\frac{1}{\left[1+r^2/(4R^2)\right]^{2+(n-1)\Gamma/2}}.
\end{equation}
Then the partition function takes the rather general form:
\begin{equation}
Z_U=\frac{K}{n!}\int d\textbf{r}_{1} \cdots \int d\textbf{r}_n \,  \prod_{i < j}\left|\det(V)\right|^{\Gamma}\prod_{i=1}^{n} g(r_i).
\label{ecu:potential_term_partition_function_VDM_determinant}
\end{equation}
In each case, $K$ and $g(r_i)$ are different.

\section{Computation for the special cases $\Gamma = 2\gamma$}

Here we briefly present the two methods used for the computation of the partition function in the cases where $\Gamma = 2\gamma$ with $\gamma$ an integer. This is, for the cases where $\Gamma$ is an even number. The first one makes use of the Jack Polynomials (see, for example, \cite{evp} and \cite{efs}), and is separated in the cases $\Gamma=2$, $\Gamma = 4p$ and $\Gamma = 4p+2$ ---with $p$ an integer---, whereas the second, due to  L. \v{S}amaj and J.K. Percus \cite{afr} maps the 2dOCP onto a one-dimensional fermionic system.

The results presented below are very simple and explicit for the case
$\Gamma=2$ and, in this case only, it is possible to obtain the
thermodynamic limit of the free energy of the plasma when
$n\to\infty$~\cite{er2d,oc2d}. For $\Gamma\geq 4$, the expression
obtained for the partition function is rather involved. It can be
computed explicitly for a small number of particles, but obtaining the
thermodynamic limit is still an open problem.

\subsection{The first method}\label{subs:first_method}  

\subsubsection{The case $\Gamma = 2$}

This first result was obtained by Jancovici back in 1980 \cite{er2d},
based on previous work by Ginibre on random matrices~\cite{sec}. In this case, (\ref{ecu:potential_term_partition_function_VDM_determinant}) becomes:
\begin{equation}
Z_U = \frac{K}{n!}\int d\mathbf{r}_{1} \cdots \int d\mathbf{r}_n \, \left|\det(V)\right|^{2} \prod_{i=1}^{n} g(r_i) = \frac{K}{n!}\int d\mathbf{r}_{1} \cdots \int d\mathbf{r}_n \, \left|\det(\Psi)\right|^{2}, 
\label{ecu:potential_term_partition_function_gamma=2}
\end{equation}
where $[\Psi]_{i,j}=\psi_{i-1}(\mathbf{r}_j)$, with
\begin{equation}
\psi_{m}(\mathbf{r}) = z^{m}\left(g(r)\right)^{1/2}.
\end{equation}

It is easily seen that this set of functions is an orthogonal set within the domain of the problem. Effectively:
\begin{equation}
\int dr \, r \int_{0}^{2\pi} d\theta \, \psi_m(\mathbf{r})\overline{\psi}_{\ell}(\mathbf{r}) = \int dr \, r^{m+\ell + 1}g(r)\int_{0}^{2\pi} d\theta \, e^{i\left(m-\ell\right)\theta} 
\end{equation} 
and the integral over $\theta$ gives zero for $\ell \neq m$, since the function $g(r)$ only depends on the magnitude $r$. Now:
\begin{equation}
\left| \det(\Psi) \right|^2 = \det \left(\Psi \right) \overline{\det}\left(\Psi\right) = \det\left(\Psi\right)\det\left( \,\overline{\Psi}\,\right).
\end{equation}
If we consider a plasma with $n=2$ ---just for illustration--- the last expression becomes:
\begin{eqnarray*}
\left| \det(\Psi) \right|^2 = \left|\psi_{0}(\mathbf{r}_1)\right|^2\left|\psi_{1}(\mathbf{r}_2)\right|^2 &+&\left|\psi_{1}(\mathbf{r}_1)\right|^2\left|\psi_{0}(\mathbf{r}_2)\right|^2\,-\\
&-&\psi_{0}(\mathbf{r}_1)\psi_{1}(\mathbf{r}_2)\overline{\psi}_{1}(\mathbf{r}_1)\overline{\psi}_{0}(\mathbf{r}_2)\,-\,\psi_{1}(\mathbf{r}_1)\psi_{0}(\mathbf{r}_2)\overline{\psi}_{0}(\mathbf{r}_1)\overline{\psi}_{1}(\mathbf{r}_2)
\end{eqnarray*}
so, after the integration, only the first two terms yield a result different from zero; actually, both results are the same, since the integration limits for each variable are equal. If we now let $n$ to take any value, the total number of integrals of this type is the total of possible permutations among the indexes of the particles. It is $n!$, and the expansion now yields:
\begin{equation}
\frac{K}{n!}\int d\mathbf{r}_{1} \cdots \int d\mathbf{r}_n \, \left| \det(\Psi) \right|^2 = K\prod_{k=0}^{n-1} \int d\mathbf{r} \left| \psi_{k}\left(\mathbf{r}\right)\right|^2.
\end{equation}
The problem reduces to calculate the norm of the functions $\psi_{k}(\mathbf{r})$. Since the norm is independent of the angle, the integration over this variable is trivial. We are left to calculate
\begin{equation}
\int dr \, r^{2k+1}g(r).
\end{equation}
For the disk, applying the change of variable $t = n r^2/R^2$, the integral above runs from zero to $n$ and becomes:
\begin{equation}
\frac{(nR^2)^{k+1}}{2}\int_{0}^{n} dt \, t^{k}e^{-t} = \frac{(nR^2)^{k+1}}{2} \gamma(k+1,n),
\end{equation}
where $\gamma(s,x)$ is the incomplete Gamma function, defined as:
\begin{equation}
\gamma(s,x) \equiv \int_{0}^{x} dt \, t^{s-1}e^{-t}
\label{ecu:incomplete_gamma_function}
\end{equation}
The partition function of the system ---its potential part--- is then
\begin{equation}
Z_U = K \pi^n \prod_{k=0}^{n-1} (nR^2)^{k+1} \gamma(k+1,n).
\label{ecu:potential_term_partition_function_gamma=2_final_soft_disk}
\end{equation}

In the soft disk case, the upper limit of the integral goes to infinity, and we obtain a complete Gamma function:
\begin{equation}
Z_U = K \pi^n \prod_{k=0}^{n-1} (nR^2)^{k+1} \Gamma(k+1) =  K \pi^n
\prod_{k=0}^{n-1} (nR^2)^{k+1} k!
\label{ecu:potential_term_partition_function_gamma=2_final_hard_disk}
\end{equation}
since $k$ is a positive integer. Finally, for the sphere:
\begin{equation}
Z_U = K \pi^n \prod_{k=0}^{n-1}\int_{0}^{\infty}dr \, \frac{r^{2k+1}}{\left[1+r^2/(4R^2)\right]^{n+1}}
\end{equation}
and, using the property
\begin{equation}
\int_0^{\infty}\frac{r^p}{(1+r)^q}dr = \frac{\Gamma(p+1)\Gamma(q-p-1)}{\Gamma(q)} = \frac{p!(q-p-2)!}{(q-1)!}
\end{equation}
for $p$ and $q$ positive integers and $q > p+1$, we have
\begin{equation}
Z_U = K\pi^n \prod_{k=0}^{n-1} 2^{2k+1}R^{2(k+1)}\frac{k!(n-k-1)!}{n!}.
\end{equation}

\subsubsection{The case $\Gamma = 4p$}

The partition function is:
\begin{equation}
Z_U = \frac{K}{n!}\int d\mathbf{r}_{1} \cdots \int d\mathbf{r}_n \, \left|\det(V)\right|^{4p} \prod_{i=1}^{n} g(r).
\label{ecu:partition_function_for_Gamma_4p}
\end{equation}

Now
\begin{equation}
\left|\det(V)\right|^{4p}=\det(V)^{2p}\overline{\det}(V)^{2p}=\prod_{i<j}(z_i-z_j)^{2p}\prod_{i<j}(\overline{z}_i-\overline{z}_j)^{2p},
\end{equation}
where the indexes run from 1 to $n$. Expanding the first term we get a symmetric polynomial of the form:
\begin{equation}
\prod_{i<j}(z_i-z_j)^{2p} = \sum_{\mu}c_\mu(2p)m_{\mu}(z_1,...,z_n), 
\label{ecu:product_Gamma_even}
\end{equation}  
where $\mu=(\mu_1,...,\mu_n)$ is an ordered partition of the number $n(n-1)p$ such that\footnote{Effectively, since the multiplication runs from 1 to $n$ on non-repeated indexes, there are $n(n-1)/2$ terms in it. Because each term is elevated to the power $2p$, the total order of each term in the expansion is $2p$ times $n(n-1)/2$, which is $n(n-1)p$.}
\begin{equation}
2p(n-1)\geq \mu_1 \geq ... \geq \mu_n\geq 0
\label{ecu:condition_over_partition}
\end{equation}
and
\begin{equation}
m_\mu(z_1,...,z_n)  = \frac{1}{\prod_i \lambda_i!}\sum_{\sigma \in S_n} z_1^{\sigma(\mu_1)} \cdots z_n^{\sigma(\mu_n)}\,,
\label{ecu:mu_part_monomial_symetric_function}
\end{equation}
where $S_n$ is the group of permutation of $n$ elements.

Condition (\ref{ecu:condition_over_partition}) arises from the fact that each $z_i$ interacts with, at most, $n-1$ particles, so it appears in the product $n-1$ times. The maximum power of $z_i$ is then $2p$ times $n-1$, which is precisely  what (\ref{ecu:condition_over_partition}) states. The numbers $\lambda_i$ are included for convenience, and denote the frequency of the integer $i$ in the partition\footnote{If, for example, we are dealing with two particles and $\Gamma = 4$, $n(n-1)p=4$, and a possible partition may be $\mu=(2,2)$. The monomial would be then:
$$\sum_{\sigma \in S_2} z_1^{\sigma(\mu_1)} z_2^{\sigma(\mu_2)}=z_1^2z_2^2+z_2^2z_1^2=2z_1^2z_2^2.$$
The term with the product appears in
(\ref{ecu:mu_part_monomial_symetric_function}) to compensate the
factor of $2$ (which is precisely the frequency, $\lambda_2=2$, of apparition of the
number $2$ in the partition $(2,2)$).} $\mu$.

It is now easily seen that the $m_{\mu}$ are orthogonal within the
space of integration; effectively, let $\mu$ and $\nu$ be two
partitions and $\{\lambda_i\}$, $\{\kappa_i\}$, their corresponding
set of frequencies. We have
\begin{eqnarray}
&&\int dr_1\, r_1 \cdots \int dr_n\, r_n \int_0^{2\pi}d\theta_1 \ \cdots \int_0^{2\pi}d\theta_n \, m_{\mu}(z_1,...,z_n) \overline{m_{\nu}(z_1,...,z_n)} \notag \\ &&=\frac{1}{\prod_i \lambda_i! \prod_j \kappa_j!}\sum_{\sigma \in S_n}\sum_{\xi \in S_n}\int dr_1 \, r_1^{\sigma(\mu_1)+\xi(\nu_1)+1} \cdots \int dr_n \, r_n^{\sigma(\mu_n)+\xi(\nu_n)+1} \notag \\&& \hspace{4cm} \times \int_0^{2\pi}d\theta_1 \, e^{i(\sigma(\mu_1)-\xi(\nu_1))\theta_1} \cdots \int_0^{2\pi}d\theta_n \, e^{i(\sigma(\mu_n)-\xi(\nu_n))\theta_n},
\label{ecu:internal_product_angle_monomial}
\end{eqnarray}
which is zero if $\sigma(\mu_i)\neq \xi(\nu_i)$ for any $1 \leq i \leq n$. This implies that the two partitions must be the same\footnote{This would not be true if not were for the fact that $\mu$ and $\nu$ are ordered partitions.}. Hence the partition function becomes
\begin{eqnarray}
Z_U&=&\frac{K}{n!}\sum_{\mu}\sum_{\nu}\frac{c_{\mu}\overline{c}_{\nu}}{\prod_i \lambda_i! \prod_j \kappa_j!}\sum_{\sigma \in S_n}\sum_{\xi \in S_n}\int dr_1 \, r_1^{\sigma(\mu_1)+\xi(\nu_1)+1}g(r_1) \cdots \notag \\ &&\times \int dr_n \, r_n^{\sigma(\mu_n)+\xi(\nu_n)+1} g(r_n) \int_0^{2\pi}d\theta_1 \, e^{i(\sigma(\mu_1)-\xi(\nu_1))\theta_1} \cdots \int_0^{2\pi}d\theta_n \, e^{i(\sigma(\mu_n)-\xi(\nu_n))\theta_n}
\end{eqnarray}
and using the orthogonality property:
\begin{equation}
Z_U = \frac{K}{n!}(2\pi)^n\sum_{\mu}\frac{\left|c_{\mu}\right|^2}{\prod_i \lambda_i!}\sum_{\sigma \in S_n}\int dr_1 \, r_1^{2\sigma(\mu_1)+1}g(r_1) \cdots \int dr_n \, r_n^{2\sigma(\mu_n)+1}g(r_n).
\end{equation}

One of the products disappears since there are $\prod_i \lambda_i!$ permutations of indexes in one of the sums that yield the same answer. Note that the permutation of powers does not change the final result of the integrals since all limits are the same. Then we have the integral of one single partition, multiplied by the ways this partition can be divided among the $r_i$ variables. This number is, of course, $n!$, and then
\begin{equation}
Z_U = K(2\pi)^n \sum_{\mu}\frac{\left|c_{\mu}\right|^2}{\prod_i \lambda_i!}\int dr_1 \, r_1^{2\mu_1+1}g(r_1) \cdots \int dr_n \, r_n^{2\sigma\mu_n+1}g(r_n).
\end{equation}
Let
\begin{equation}
G_{\mu_{\ell}} = 2 \int dr \, r^{2\mu_{\ell}+1}g(r)
\end{equation}
so, finally:
\begin{equation}
Z_U = K \pi^n \sum_{\mu}\frac{\left|c_{\mu}\right|^2}{\prod_i \lambda_i!}\prod_{\ell = 1}^n G_{\mu_{\ell}}.
\label{ecu:general_expression_partition_function_gamma_4p}
\end{equation}
The $c_{\mu}(2p)$ can be obtained from (\ref{ecu:mu_part_monomial_symetric_function}) using the property of orthogonality:
\begin{equation}
c_{\mu}(2p)=\int_0^{2\pi} d\theta_1 \, e^{-i\mu_1\theta_1} \cdots \int_0^{2\pi} d\theta_n \,  e^{-i\mu_n\theta_n} \prod_{j < k}\left(e^{i\theta_j}-e^{i\theta_k}\right)^{2p}.
\end{equation}

\subsubsection{The case $\Gamma = 4p+2$}
Here, the partition function is similar to (\ref{ecu:partition_function_for_Gamma_4p}) changing $2p$ for $2p+1$:
\begin{equation}
\frac{K}{n!}\int d\mathbf{r}_{1} \cdots \int d\mathbf{r}_n \, \left|\det(V)\right|^{4p+2} \prod_{i=1}^{n} g(r_i).
\end{equation} 
The product of differences is now raised to an odd power
\begin{equation}
\left|\det(V)\right|^{4p+2} = \det(V)^{2p+1}\overline{\det}(V)^{2p+1}=\prod_{i<j}(z_i-z_j)^{2p+1}\prod_{i<j}(\overline{z}_i-\overline{z}_j)^{2p+1}
\end{equation}
and then the polynomials are now anti-symmetric, yielding to an expansion of the form:
\begin{equation}
\prod_{i<j}(z_i-z_j)^{2p+1} = \sum_{\mu}c_{\mu}(2p+1)q_{\mu}(z_1,...,z_n),
\label{ecu:product_Gamma_odd}
\end{equation}
where $\mu$ is now an ordered partition of the number $n(n-1)(2p+1)/2$ such that
\begin{equation}
(n-1)(2p+1) \geq \mu_1 \geq \cdots \geq \mu_n \geq 0
\end{equation}
and
\begin{equation}
q_{\mu}(z_1,\ldots,z_n)=\frac{1}{\prod_{i}\lambda_i!}\sum_{\sigma \in S_N}(-1)^{\sigma}z_{1}^{\sigma(\mu_1)}\cdots z_{n}^{\sigma(\mu_n)},
\end{equation}
where $(-1)^{\sigma}$ denotes the sign of the permutation $\mu$.

Actually, the frecuency $\lambda_i$ is always 1. This is due to the
fact that a transposition of two indexes changes a sign. However, we leave the expression here to highlight the similarities with the previous case. Again, using the orthogonality of the $q_{\mu}$ monomial functions over angular integration
\begin{eqnarray*}
Z_U&=&\frac{K}{N!}\sum_{\mu}\sum_{\nu}\frac{c_{\mu}\overline{c}_{\nu}}{\prod_i \lambda_i! \prod_j \kappa_j!}\sum_{\sigma \in S_n}\sum_{\xi \in S_n}(-1)^{\sigma+\xi}\int dr_1 \, r_1^{\sigma(\mu_1)+\xi(\nu_1)+1}g(r_1) \cdots \\ &\times& \int dr_n \, r_n^{\sigma(\mu_n)+\xi(\nu_n)+1} g(r_n) \int_0^{2\pi}d\theta_1 \, e^{i(\sigma(\mu_1)-\xi(\nu_1))\theta_1} \cdots \int_0^{2\pi}d\theta_n \, e^{i(\sigma(\mu_n)-\xi(\nu_n))\theta_n}.
\end{eqnarray*}
we obtain again $\mu = \nu$; then
\begin{equation}
Z_U = \frac{K}{n!}(2\pi)^n\sum_{\mu}\frac{\left|c_{\mu}\right|^2}{\prod_i \lambda_i!}\sum_{\sigma \in S_n}\int dr_1 \, r_1^{2\sigma(\mu_1)+1}g(r_1) \cdots \int dr_n \, r_n^{2\sigma(\mu_n)+1}g(r_n).
\end{equation}
Note that the sign disappears since the permutations $\sigma$ and $\xi$ must be the same. Defining:
\begin{equation}
G_{\mu_{\ell}}= 2\int dr \, r^{2\mu_{\ell}+1}g(r)
\end{equation}
we obtain
\begin{equation}
Z_U = K(\pi)^n\sum_{\mu}\frac{\left|c_{\mu}\right|^2}{\prod_i \lambda_i!}\prod_{\ell=1}^n G_{\mu_{\ell}}.
\label{ecu:general_expression_partition_function_gamma_4p+2}
\end{equation}
The coefficients $c_{\mu}(2p+1)$ can be calculated in the same way as were the $c_{\mu}(2p)$:
\begin{equation}
c_{\mu}(2p+1)=\int_0^{2\pi} d\theta_1 \, e^{-i\mu_1\theta_1} \cdots \int_0^{2\pi} d\theta_n \,  e^{-i\mu_n\theta_n} \prod_{j < k}\left(e^{i\theta_j}-e^{i\theta_k}\right)^{2p+1}.
\end{equation}

This completes the computation of the partition function by the means
of the symmetric or antisymmetric monomials ---the $m_{\mu}$ and $q_{\mu}$ functions---. Now we proceed with the mapping onto the one-dimensional fermion's chain.

\subsection{The second method}\label{subs:second_method}

This method, proposed by L. \v{S}amaj and J.K. Percus \cite{afr} maps the 2dOCP onto a one-dimensional fermionic system. They use the following property of Grassmann quadratic forms:
\begin{equation}
\int_{\Lambda} \prod_{i=1}^n d\eta_i d\overline{\eta}_i \, e^{\overline{\eta} A\eta} = \det{A},
\end{equation}
where $\eta_i$ and $\overline{\eta}_i$ are Grassmann variables. With this, it is possible to rewrite the partition function of the plasma (equation \ref{ecu:potential_term_partition_function_VDM_determinant}) with $\Gamma = 2\gamma$ as
\begin{eqnarray}
Z_U &&= \frac{K}{n!}\int d\mathbf{r}_{1} \cdots \int d\mathbf{r}_n \, \left|\det(V)\right|^{2\gamma} \prod_{i=1}^{n} g(r_i) \notag \\
 &&= \frac{K}{n!}\int d\mathbf{r}_{1} \cdots \int d\mathbf{r}_n \, \det \,_{(1)}(V) \cdots \det \,_{(\gamma)}(V) \det \,_{(1)}(\overline{V})\cdots \det \,_{(\gamma)}(\overline{V})  \prod_{i=1}^{n} g(r_i) \notag 
\end{eqnarray} 
and replace each determinant as a Gaussian integral over $2n$ Grassmann variables, say $(\xi^{(\alpha)}_k,\overline{\xi}^{(\alpha)}_k)$ and $(\psi^{(\alpha)}_k,\overline{\psi}^{(\alpha)}_k)$, with $1 \leq \alpha \leq \gamma$ and $1 \leq k \leq n$. Then
\begin{eqnarray}
&\det \, _{(\alpha)}(V) = \int \prod_{k=1}^{n} d\xi_{k}^{(\alpha)}d \overline{\xi}_{k}^{(\alpha)} e^{\overline{\xi}^{(\alpha)} V \xi^{(\alpha)}} = \int \prod_{k=1}^{n} d\xi_{k}^{(\alpha)}d \overline{\xi}_{k}^{(\alpha)} \left(1 + \overline{\xi}^{(\alpha)}_k\sum_{j} \left[V\right]_{k,j}\xi_j^{(\alpha)}\right)&\\
&\det \, _{(\alpha)}(\overline{V}) = \int \prod_{k=1}^{n}
  d\psi_{k}^{(\alpha)}d \overline{\psi}_{k}^{(\alpha)}
  e^{\overline{\psi}^{(\alpha)} \overline{V} \psi^{(\alpha)}} = \int
  \prod_{k=1}^{n} d\psi_{k}^{(\alpha)}d \overline{\psi}_{k}^{(\alpha)}
  \left(1 + \overline{\psi}_k^{(\alpha)}\sum_{j}
  \left[\overline{V}\right]_{k,j}\psi_j^{(\alpha)}\right)
\,.
&
\end{eqnarray}
In the previous equations, we use the fact that $e^a=1+a$, for any
Grassmann variable $a$, because $a^{2}=0$ due to the anticomutation
rules. So the partition function then becomes
\begin{eqnarray}
Z_U =  \frac{K}{n!}\int d\mathbf{r}_{1} \cdots \int d\mathbf{r}_n \, \int_{\Lambda} \prod_{k=1}^{n} \left. \Bigg( d\xi^{(1)}_kd\overline{\xi}^{(1)}_k \cdots d\xi^{(\gamma)}_kd\overline{\xi}^{(\gamma)}_k \, d\psi_{k}^{(1)}d \overline{\psi}_{k}^{(1)} \cdots d\psi_{k}^{(\gamma)}d \overline{\psi}_{k}^{(\gamma)} \right. \times \notag \\
\left. \times \prod_{\alpha=1}^{\gamma}\left[1 + \overline{\xi}^{(\alpha)}_k \sum_{j} \left[V\right]_{k,j}\xi_j^{(\alpha)}\right]\left[1 + \overline{\psi}_k^{(\alpha)} \sum_{m} \left[\overline{V}\right]_{k,m}\psi_m^{(\alpha)}\right]\right)\prod_{\ell=1}^n g(r_{\ell}).
\end{eqnarray}

Since the only term which is nonzero in the integral is that in which all the variables appear, we may as well replace the product of the two terms in square brackets by
\begin{eqnarray}
&&\prod_{\alpha=1}^{\gamma}\left[1 + \overline{\xi}^{(\alpha)}_k \sum_{j} \left[V\right]_{k,j}\xi_j^{(\alpha)}\right]\left[1 + \overline{\psi}_k^{(\alpha)} \sum_{m} \left[\overline{V}\right]_{k,m}\psi_m^{(\alpha)}\right] \notag \\
&&\longrightarrow \overline{\psi}_k^{(\gamma)}\cdots\overline{\psi}_k^{(1)}\overline{\xi}_k^{(\gamma)}\cdots\overline{\xi}_k^{(1)} \sum_{j_1,...,j_{\gamma}=1}^{n}\sum_{m_1,...,m_{\gamma}=1}^{n}\xi_{j_1}^{(1)}\cdots\xi_{j_{\gamma}}^{(\gamma)}r_k^{J+M-2\gamma}e^{i\theta_{k}(J-M)}\psi^{(1)}_{m_1}\cdots\psi^{( \gamma)}_{m_{\gamma}},
\end{eqnarray}
where $J = j_1 + \cdots + j_{\gamma}$ and $M = m_1 + \cdots + m_{\gamma}$ and we have replaced $\left[V\right]_{k,j}=\left(r_ke^{i\theta_k}\right)^{j-1}$. The integration over the conjugated variables is straightforward and yields $1$, leaving
\begin{eqnarray}
&&Z_U = \frac{K}{n!}\int d\mathbf{r}_{1} \cdots \int d\mathbf{r}_n \, \int_{\Lambda}\prod_{k=1}^{n}\left. \Bigg( d\xi_{k}^{(1)} \cdots d\xi_{k}^{(\gamma)} d\psi_{k}^{(1)} \cdots d\psi_{k}^{(\gamma)} \times\right. \notag \\
&& \hspace{1.5cm} \left.\times \sum_{j_1,...,j_{\gamma}=1}^{n}\sum_{m_1,...,m_{\gamma}=1}^{n}\xi_{j_1}^{(1)}\cdots\xi_{j_{\gamma}}^{(\gamma)}r_k^{J+M-2\gamma}e^{i\theta_{k}(J-M)}\psi^{(1)}_{m_1}\cdots\psi^{( \gamma)}_{m_{\gamma}}\right)\prod_{\ell=1}^n g(r_{\ell}).
\end{eqnarray}

Note now that the factor
\begin{eqnarray}
\sum_{j_1,...,j_{\gamma}=1}^{n}\sum_{m_1,...,m_{\gamma}=1}^{n}\xi_{j_1}^{(1)}\cdots\xi_{j_{\gamma}}^{(\gamma)}\int d\mathbf{r} \, r^{J+M-2\gamma}e^{i\theta(J-M)}g(r)\psi^{(1)}_{m_1}\cdots\psi^{( \gamma)}_{m_{\gamma}}
\end{eqnarray}
occurs $n$ times in the last expression, yielding
\begin{eqnarray}
&& Z_U = \frac{K}{n!} \int_{\Lambda}\prod_{k=1}^n  \left(d\xi_{k}^{(1)} \cdots d\xi_{k}^{(\gamma)} d\psi_{k}^{(1)} \cdots d\psi_{k}^{(\gamma)}\right) \times \notag \\
&& \hspace{1cm} \times \left(\sum_{j_1,...,j_{\gamma}=1}^{n}\sum_{m_1,...,m_{\gamma}=1}^{n}\xi_{j_1}^{(1)}\cdots\xi_{j_{\gamma}}^{(\gamma)}\int d\mathbf{r} \, r^{J+M-2\gamma}e^{i\theta(J-M)} g(r)\psi^{(1)}_{m_1}\cdots\psi^{( \gamma)}_{m_{\gamma}}\right)^n.
\label{ecu:partition_function_method_fermions}
\end{eqnarray}

At first sight, it may seem that this result is more complicated than the results from the previous section. However, as we will see next, this expression allows an easy form to calculate the coefficients $c_{\mu}$ of the equations (\ref{ecu:general_expression_partition_function_gamma_4p}) and (\ref{ecu:general_expression_partition_function_gamma_4p+2}), instead of the complex integrals proposed.

\section{The merging}

First, it is convenient to sum up the results until now; from the first treatment done in Section ~\ref{subs:first_method} we have:
\begin{eqnarray}
&&Z_U = K \pi^n \sum_{\mu}\frac{\left|c_{\mu}\right|^2}{\prod_i \lambda_i!}\prod_{\ell = 1}^n G_{\mu_{\ell}}\label{ecu:partition_meth_1_general}\\
&&G_{\mu_{\ell}} =  2 \int dr \, r^{2\mu_{\ell}+1}g(r)\label{ecu:G_general}\\
&&c_{\mu}=\int_0^{2\pi} d\theta_1 \, e^{-i\mu_1\theta_1} \cdots \int_0^{2\pi} d\theta_n \,  e^{-i\mu_n\theta_n} \prod_{j < k}\left(e^{i\theta_j}-e^{i\theta_k}\right)^{\gamma}\label{ecu:C_general}
\end{eqnarray} 
where $\mu=(\mu_1,...,\mu_n)$ is an ordered partition of the number $\gamma n(n-1)/2$, with $\gamma = \Gamma/2$ such that
\begin{equation}
\gamma (n-1) \geq \mu_1 \geq ... \geq \mu_n \geq 0
\label{ecu:partition_general}
\end{equation}
and $\lambda_i$ is the times the number $i$ is repeated in the partition. Now, from the second method in Section ~\ref{subs:second_method} we have:
\begin{eqnarray}
&& Z_U = \frac{K}{n!} \int_{\Lambda}\prod_{k=1}^n
  \left(d\xi_{k}^{(1)} \cdots d\xi_{k}^{(\gamma)} d\psi_{k}^{(1)}
  \cdots d\psi_{k}^{(\gamma)}\right) \times \notag 
\nonumber\\
&& \hspace{1cm} \times \left(\sum_{j_1,...,j_{\gamma}=1}^{n}\sum_{m_1,...,m_{\gamma}=1}^{n}\xi_{j_1}^{(1)}\cdots\xi_{j_{\gamma}}^{(\gamma)}\int d\mathbf{r} \, r^{J+M-2\gamma}e^{i\theta(J-M)} g(r)\psi^{(1)}_{m_1}\cdots\psi^{( \gamma)}_{m_{\gamma}}\right)^n.
\end{eqnarray}

We now want to find a way to get the results of the first treatment from this last expression. In order to do this, first note that the spatial integration is zero unless $J=M$, because of the complex exponential. In this case, the angular integral is $2\pi$. Because of this, the partition function can be rewritten as:
\begin{eqnarray}
&& Z_U = \frac{K\pi^n}{n!} \int_{\Lambda}\prod_{k=1}^n  \left(d\xi_{k}^{(1)} \cdots d\xi_{k}^{(\gamma)} d\psi_{k}^{(1)} \cdots d\psi_{k}^{(\gamma)}\right) \times \notag \\
&& \hspace{1cm} \times \left(\sum_{j_1,...,j_{\gamma}=1}^{n}\sum_{m_1,...,m_{\gamma}=1}^n\xi_{j_1}^{(1)}\cdots\xi_{j_{\gamma}}^{(\gamma)}2\int dr \, r^{2(J-\gamma)+1}g(r)\psi^{(1)}_{m_1}\cdots\psi^{( \gamma)}_{m_{\gamma}}\delta_{j_1+\cdots + j_{\gamma},J}\delta_{m_1+\cdots + m_{\gamma},J}\right)^n,
\end{eqnarray}
where the Kronecker deltas ensure the condition:
\begin{equation}
\sum_{i=1}^{\gamma}m_i = \sum_{i=1}^{\gamma}j_i = J \notag
\end{equation}
Now, the term in brackets can be rewritten as:
\begin{eqnarray}
&&\left(\sum_{j_1,...,j_{\gamma}=1}^{n}\sum_{m_1,...,m_{\gamma}=1}^n\xi_{j_1}^{(1)}\cdots\xi_{j_{\gamma}}^{(\gamma)}2\int dr \, r^{2(J-\gamma)+1}g(r)\psi^{(1)}_{m_1}\cdots\psi^{( \gamma)}_{m_{\gamma}}\right)^n  \notag \\
&& \hspace{1cm} = \prod_{i=1}^{n}\sum_{j^{(i)}_1,...,j^{(i)}_{\gamma}=1}^{n}\sum_{m^{(i)}_1,...,m^{(i)}_{\gamma}=1}^{n}\xi_{j^{(i)}_1}^{(1)}\cdots\xi_{j^{(i)}_{\gamma}}^{(\gamma)}\psi^{(1)}_{m^{(i)}_1}\cdots\psi^{( \gamma)}_{m^{(i)}_{\gamma}}\delta_{j_1^{(i)}+\cdots + j_{\gamma}^{(i)},J^{(i)}}\delta_{m_1^{(i)}+\cdots + m_{\gamma}^{(i)},J^{(i)}}G_{J^{(i)}-\gamma},
\end{eqnarray}
where the expression $G_{J^{(i)}-\gamma}$ is defined in (\ref{ecu:G_general}). In the Grassmann integral, the only terms that survive are those which include all the Grassmann variables only once. This implies that we require $j^{(i)}_{\ell} \neq j^{(m)}_{\ell}$ for $i \neq m$. Now
\begin{equation}
J^{(i)}=\sum_{\ell=1}^{\gamma} j^{(i)}_{\ell} = \sum_{\ell=1}^{\gamma} m^{(i)}_{\ell}
\end{equation}
so
\begin{equation}
\sum_{i=1}^{n}J^{(i)} = \sum_{i=1}^{n}\sum_{\ell=1}^{\gamma} j^{(i)}_{\ell} = \sum_{\ell=1}^{\gamma}\sum_{i=1}^{n}j^{(i)}_{\ell},
\end{equation}
but since $j^{(i)}_{\ell}$ runs from $1$ to $n$ and for a given $\ell$ we have $j^{(i)}_{\ell} \neq j^{(m)}_{\ell}$, the first sum is just
\begin{equation}
\sum_{i=1}^{n}j^{(i)}_{\ell} = \sum_{i=1}^{n}i = \frac{n(n+1)}{2}
\end{equation}
and then
\begin{equation}
\sum_{i=1}^{n}J^{(i)} = \frac{\gamma}{2}n(n+1).
\end{equation}

We can take the different $J^{(i)}$ as partitions of the number $\gamma n(n+1)/2$. Since the minimum of each $j^{(i)}_{\ell}$ is $1$ and its maximum is $n$, we see that each $J^{(i)}$ must fulfill the requirement
\begin{equation}
\gamma n \geq J^{(i)} \geq \gamma.
\end{equation}

We define $\mathcal{J}=\left(J^{(1)},...,J^{(n)}\right)$ as the ordered partition of the number $\gamma n(n+1)/2$ such that
\begin{equation}
\gamma n \geq J^{(1)} \geq ... \geq J^{(n)} \geq \gamma.
\end{equation}
This is similar to the expression in (\ref{ecu:partition_general}); effectively, it is easily seen that $\mu_i = J^{(i)}-\gamma$. Let
\begin{eqnarray}
&&\Phi({J^{(i)}}) = \sum_{j^{(i)}_1,...,j^{(i)}_{\gamma}=1}^{n}\sum_{m^{(i)}_1,...,m^{(i)}_{\gamma}=1}^{n}\xi_{j^{(i)}_1}^{(1)}\cdots\xi_{j^{(i)}_{\gamma}}^{(\gamma)}\psi^{(1)}_{m^{(i)}_1}\cdots\psi^{( \gamma)}_{m^{(i)}_{\gamma}}\delta_{j_1^{(i)}+\cdots + j_{\gamma}^{(i)},J^{(i)}}\delta_{m_1^{(i)}+\cdots + m_{\gamma}^{(i)},J^{(i)}}.
\end{eqnarray}
The partition function then reads
\begin{eqnarray}
Z_U = \frac{K\pi^n}{n!} \int_{\Lambda}\prod_{k=1}^n  \left(d\xi_{k}^{(1)} \cdots d\xi_{k}^{(\gamma)} d\psi_{k}^{(1)} \cdots d\psi_{k}^{(\gamma)}\right)\sum_{\mathcal{J}}\frac{1}{\prod_i \lambda_i!}\sum_{\sigma \in S^n}\prod_{i=1}^{n}\Phi\left(\sigma\left({J}^{(i)}\right)\right)G_{\sigma(J^{(i)})-\gamma}
\end{eqnarray}
where $\lambda_i$ is the number of times the number $i$ appears in the $\mathcal{J}$ partition. Now, note that the function $\Phi\left(J^{(i)}\right)$ contains an even number of Grassmann variables, so it commutes with any other function. This implies that the sum over the permutation is just $n!$ times the same expression. Finally, interchanging the integration and the sum over the partition we get
\begin{equation}
Z_U = K\pi^n \sum_{\mathcal{J}}\frac{1}{\prod_i \lambda_i!}\left[\int_{\Lambda}\prod_{k=1}^n d\xi_{k}^{(1)} \cdots d\xi_{k}^{(\gamma)} d\psi_{k}^{(1)} \cdots d\psi_{k}^{(\gamma)}\Phi\left(J^{(k)}\right)\right]\prod_{\ell=1}^{n}G_{J^{(\ell)}-\gamma}
\end{equation}
but $J^{(\ell)}-\gamma = \mu_{\ell}$, so $\mathcal{J}$ can be replaced by a partition $\mu$ of the number $\gamma n(n+1)/2 - n\gamma = \gamma n(n-1)/2$ such that
\begin{equation*}
\gamma n-\gamma = \gamma(n-1) \geq \mu_1 \geq ... \geq \mu_n \geq \gamma - \gamma = 0
\end{equation*}
so
\begin{equation}
Z_U = K\pi^n \sum_{\mu}\frac{1}{\prod_i \lambda_i!}\left[\int_{\Lambda}\prod_{k=1}^n d\xi_{k}^{(1)} \cdots d\xi_{k}^{(\gamma)} d\psi_{k}^{(1)} \cdots d\psi_{k}^{(\gamma)}\Phi\left(\mu_k+\gamma\right)\right]\prod_{\ell=1}^{n}G_{\mu_{\ell}}
\end{equation}
and we recover (\ref{ecu:partition_meth_1_general}), provided that
\begin{equation}
\int_{\Lambda}\prod_{k=1}^n d\xi_{k}^{(1)} \cdots d\xi_{k}^{(\gamma)} d\psi_{k}^{(1)} \cdots d\psi_{k}^{(\gamma)}\Phi\left(\mu_k+\gamma\right) = \left|c_{\mu}\right|^2.
\end{equation}

There is a simpler way to express the relation stated above for the coefficients $c_{\mu}$. Since
\begin{equation}
\Phi(J^{(i)})=\Xi(J^{(i)};\xi)\,\Xi(J^{(i)};\psi)
\end{equation}
with
\begin{equation}
\Xi(J^{(i)};\xi) =  \sum_{j^{(i)}_1,...,j^{(i)}_{\gamma}=1}^{n}\xi_{j^{(i)}_1}^{(1)}\cdots\xi_{j^{(i)}_{\gamma}}^{(\gamma)}\delta_{j_1^{(i)}+\cdots + j_{\gamma}^{(i)},J^{(i)}}
\end{equation}
we have
\begin{eqnarray}
&& \int_{\Lambda}\prod_{k=1}^n d\xi_{k}^{(1)} \cdots d\xi_{k}^{(\gamma)} d\psi_{k}^{(1)} \cdots d\psi_{k}^{(\gamma)}\Phi\left(\mu_k+\gamma\right)\notag \\
&& = \int_{\Lambda}\prod_{k=1}^n d\xi_{k}^{(1)} \cdots d\xi_{k}^{(\gamma)} d\psi_{k}^{(1)} \cdots d\psi_{k}^{(\gamma)}\Xi\left(\mu_k+\gamma;\xi\right)\Xi\left(\mu_k+\gamma;\psi\right). 
\end{eqnarray}

Permuting:
\begin{eqnarray}
&& \int_{\Lambda}\prod_{k=1}^n d\xi_{k}^{(1)} \cdots d\xi_{k}^{(\gamma)} d\psi_{k}^{(1)} \cdots d\psi_{k}^{(\gamma)}\Xi\left(\mu_k+\gamma;\xi\right)\Xi\left(\mu_k+\gamma;\psi\right) \\
&& = \int_{\Lambda}\prod_{k=1}^n d\xi_{k}^{(1)} \cdots d\xi_{k}^{(\gamma)}\Xi\left(\mu_k+\gamma;\xi\right)\int_{\Lambda}\prod_{k=1}^n d\psi_{k}^{(1)} \cdots d\psi_{k}^{(\gamma)}\Xi\left(\mu_k+\gamma;\psi\right) \notag \\
&& = \left(\int_{\Lambda} \prod_{k=1}^{n} d\xi_k^{(1)}\cdots d\xi_{k}^{(\gamma)}\Xi\left(\mu_k+\gamma;\xi\right)\right)^2.
\end{eqnarray}

So, up to a (possible) phase factor:
\begin{equation}
c_{\mu} = \int_{\Lambda} \prod_{k=1}^{n} d\xi_k^{(1)}\cdots d\xi_{k}^{(\gamma)}\Xi\left(\mu_k+\gamma;\xi\right).
\label{ecu:expression_coefficients_grassmann_integral}
\end{equation}

We gain another expression for the expansion coefficients; however, the computation of these is still troublesome. Because of this, in the next section we derive a series of rules which intend to facilitate their calculation. 

\section{Properties of the expansion coefficients}

\subsection{Recurrence relation for the coefficients obtained from the
Grassmann representation}

The coefficients for the expansion can be calculated directly from the expression in (\ref{ecu:expression_coefficients_grassmann_integral}). However,
\v{S}amaj \cite{i2ds} has found a recurrence relation derived from
this equation, which we review below. The function $\Xi(J^{(i)})$ can be rewritten as:
\begin{equation}
 \Xi(J^{(i)})=\sum_{j_1^{(i)},...,j_{\gamma}^{(i)}=1}^{n}\xi^{(1)}_{j_1^{(i)}}\cdots\xi^{(\gamma)}_{j_{\gamma}^{(i)}} \delta_{j_1^{(i)}+\cdots+j_{\gamma}^{(i)},J^{(i)}}.
 \label{ecu:expression_coefficients_sum_delta}
\end{equation}
On the other hand, it is evident that
\begin{equation}
 c_{\gamma(n-1),\gamma(n-2),...,0}=\int_{\Lambda} \prod_{k=1}^{n} d\xi_k^{(1)}\cdots d\xi_{k}^{(\gamma)}\Xi\left(k\gamma;\xi\right).
\end{equation}
For $k=1$, the only possible values for the coefficients $j_1^{(i)},...,j_{\gamma}^{(i)}$ are $1$. Therefore
\begin{equation}
 \Xi\left(\gamma;\xi\right)=\xi^{(1)}_{1}\cdots\xi^{(\gamma)}_{1}.
\end{equation}
For $k=2$, the only possible values for these coefficients are $2$ (there cannot be $3$ since this would imply for another coefficient to be $1$,
which is forbidden since these values are already taken by $\Xi(\gamma)$). Then
\begin{equation}
  \Xi\left(2\gamma;\xi\right)\Xi\left(\gamma;\xi\right)=\xi^{(1)}_{2}\cdots\xi^{(\gamma)}_{2}\xi^{(1)}_{1}\cdots\xi^{(\gamma)}_{1}
\end{equation}
It is easily seen then that the product is
\begin{equation}
 \prod_{k=1}^{n} d\xi_k^{(1)}\cdots d\xi_{k}^{(\gamma)}\Xi\left(k\gamma;\xi\right) = \prod_{k=1}^{n} d\xi_k^{(1)}\cdots d\xi_{k}^{(\gamma)}\xi_{k}^{(1)}\cdots\xi_{k}^{(\gamma)}
\end{equation}
so
\begin{equation}
c_{\gamma(n-1),\gamma(n-2),...,0} = 1.
\label{ecu:coefficients_gamma_series}
\end{equation}
From this, we see that the partition
\begin{equation}
\kappa=(\gamma(n-1),\gamma(n-2),...,0)
\label{ecu:root_partition}
\end{equation}
plays a special role. This
partition will be called the root partition~\cite{aana,dfqh}.

Now, the parity ---even or odd--- of the number $\gamma$ will define some relations among the functions $\Xi(J^{(i)})$. If $\gamma$ is even
\begin{equation}
 \left[\Xi(J^{(i)}),\Xi(J^{(k)})\right]=0
\end{equation}
since every term in the expansion contains an even number of Grassmann variables. On the other hand, if $\gamma$ is odd
\begin{equation}
 \left\lbrace \Xi(J^{(i)}),\Xi(J^{(k)})\right\rbrace=0
\end{equation}
where $\left\lbrace,\right\rbrace$ denotes anti-commutation. This implies that any permutation $\sigma$ of the indexes of a coefficient yields
\begin{equation}
c_{\sigma(\mu_1),\sigma(\mu_2),...,\sigma(\mu_n)}=\left[(-1)^{\sigma}\right]^{\gamma}c_{\mu_1,...,\mu_n}
 \label{ecu:change_sign_permutation_indexes}
\end{equation}
with $(-1)^{\sigma}$ the sign of the permutation. Finally, let us construct a polynomial $\mathcal{P}$ of order less or equal to $(\gamma - 1)$ in the variables $J^{(i)},J^{(k)}$. 
Then, consider this polynomial in the following combination with the functions $\Xi$: 
\begin{equation}
 \sum_{J^{(i)},J^{(k)}=\gamma}^{n\gamma} \mathcal{P}(J^{(i)},J^{(k)})\Xi(J^{(i)})\Xi(J^{(k)})\delta_{J^{(i)}+J^{(j)},J}
\end{equation}
with $J$ being any integer from $2\gamma$ to $2n\gamma$. Rewriting this in terms of the $\xi$ variables:
\begin{equation}
 \sum_{\substack{j_1^{(i)},...,j_{\gamma}^{(i)}=1\\j_1^{(k)},...,j_{\gamma}^{(i)}=1}}^{n} \mathcal{P}(j_1^{(i)}+\cdots+j_{\gamma}^{(i)},j_1^{(k)}+\cdots+j_{\gamma}^{(k)})\xi_{j_1^{(i)}}^{(i)}\cdots \xi_{j_{\gamma}^{(1)}}^{(1)}\xi_{j_1^{(k)}}^{(\gamma)}\cdots \xi_{j_{\gamma}^{(k)}}^{(\gamma)}\delta_{J,j_1^{(i)}+\cdots+j_{\gamma}^{(i)}+j_1^(k)+\cdots+j_{\gamma}^{(i)}}.
\label{ecu:poly_recurrence_relation}
\end{equation}
Since the polynomial is of order $(\gamma-1)$, each term in the expansion of $\mathcal{P}$ lacks at least one couple of the form $(j_m^{(i)},j_m^{(k)})$
---not necessarily the same for every term---. This implies that the sum in (\ref{ecu:poly_recurrence_relation}) can be separated in expressions of the form:
\begin{equation}
 A(m,i,k,J,\gamma)\sum_{j_m^{(i)},j_m^{(k)} = 1}^{n}\xi_{j_m^{(i)}}^{(m)}\xi_{j_m^{(k)}}^{(m)}\delta_{J,j_1^{(i)}+\cdots+j_{\gamma}^{(i)}+j_1^(k)+\cdots+j_{\gamma}^{(i)}}
\end{equation}
where $A(m,i,k,J,\gamma)$ contains the dependence on the other Grassmann variables and on the polynomial expansion. But
\begin{equation}
\sum_{j_m^{(i)},j_m^{(k)} = 1}^{n}\xi_{j_m^{(i)}}^{(m)}\xi_{j_m^{(k)}}^{(m)}\delta_{J,j_1^{(i)}+\cdots+j_{\gamma}^{(i)}+j_1^(k)+\cdots+j_{\gamma}^{(i)}} = \sum_{j_m^{(i)} \leq j_m^{(k)}}\left(\xi_{j_m^{(i)}}^{(m)}\xi_{j_m^{(k)}}^{(m)}+\xi_{j_m^{(k)}}^{(m)}\xi_{j_m^{(i)}}^{(m)}\right)\delta_{J,j_1^{(i)}+\cdots+j_{\gamma}^{(i)}+j_1^(k)+\cdots+j_{\gamma}^{(i)}}
\label{ecu:zeroth_sum_anticommutaion_relation}
\end{equation}
which is zero since the Grassmann variables anticommute. So every term in the expansion is zero and
\begin{equation}
 \sum_{J^{(i)},J^{(k)}=\gamma}^{n\gamma} \mathcal{P}(J^{(i)},J^{(k)})\Xi(J^{(i)})\Xi(J^{(k)})\delta_{J^{(i)}+J^{(j)},J} = 0
\end{equation}
Multipliying by other functions $\Xi$ does not change this property. Integrating over the Grassman variables we get
\begin{equation}
 \sum_{\substack{J^{(i)},J^{(k)}=\gamma\\J^{(i)}+J^{(k)}=J}}^{\gamma n} \mathcal{P}(J^{(i)},J^{(k)}) c_{i_1,...,i_{n-2},J^{(i)}-\gamma,J^{(k)}-\gamma}=0
\label{ecu:recurrence_relation_general_polynomial}
\end{equation}
for $J=2\gamma,...,2n\gamma$. \v{S}amaj proposes $\mathcal{P}(J^{(i)},J^{(k)}) = \left(J^{(i)}-J^{(k)}\right)^{x}$, where $x=0,...,\gamma-1$. However,
notice that
\begin{eqnarray}
&&\sum_{\substack{J^{(i)},J^{(k)}=\gamma\\J^{(i)}+J^{(k)}=J}}^{\gamma n} \mathcal{P}(J^{(i)},J^{(k)}) c_{i_1,...,i_{n-2},J^{(i)}-\gamma,J^{(k)}-\gamma} \notag \\
&&=\sum_{\substack{J^{(k)} < J^{(i)}\\J^{(i)}+J^{(k)}=J}}\left[ \left(J^{(i)}-J^{(k)}\right)^xc_{i_1,...,i_{n-2},J^{(i)}-\gamma,J^{(k)}-\gamma}+\left(J^{(k)}-J^{(i)}\right)^xc_{i_1,...,i_{n-2},J^{(k)}-\gamma,J^{(i)}-\gamma}\right] \notag \\
&&=\sum_{\substack{J^{(k)} < J^{(i)}\\J^{(i)}+J^{(k)}=J}}\left[ \left(J^{(i)}-J^{(k)}\right)^x+\left(-1\right)^{\gamma}\left(J^{(k)}-J^{(i)}\right)^x\right]c_{i_1,...,i_{n-2},J^{(k)}-\gamma,J^{(i)}-\gamma}.
\end{eqnarray}
If $\gamma$ is even, an odd value of $x$ would make the expression in the square braquets zero for any partition. In the same way, if $\gamma$ is
odd, an even value of $x$ would cancel this term. This relation, then, gives information on the coefficients only when the power $x$ has the same
parity than the number $\gamma$. The easiest values are $x=0$ for $\gamma$ even and $x=1$ for $\gamma$ odd. Replacing $J^{(k)}-\gamma=i_{n-1}$ and
$J^{(i)}-\gamma=i_n$, we get that, if $\gamma$ is even
\begin{eqnarray}
 \sum_{\substack{i_{n} < i_{n-1}\\i_{n-1}+i_{n}=K}}c_{i_1,...,i_{n-2},i_{n-1},i_n} = -\frac{1}{2}c_{i_1,...,i_{n-2},K/2,K/2}.
 \label{ecu:recurrence_grassmann_even}
\end{eqnarray}
If $K$ is odd, the term on the right is zero. If $\gamma$ is odd:
\begin{equation}
 \sum_{\substack{i_{n} < i_{n-1}\\i_{n-1}+i_{n}=K}}\left(i_{n-1}-i_{n}\right)c_{i_1,...,i_{n-2},i_{n-1},i_n} = 0 
 \label{ecu:recurrence_grassmann_odd}
\end{equation}
with $K=0,...,2\gamma(n-1)$. Despite the fact the sum seems to be done
over the last two components of the partition, it is equally valid
with any choice of pairs ---this would have implied multiplying by
other $\Xi$ function on the left, on the right and in between in
(\ref{ecu:recurrence_relation_general_polynomial})---. However, the
$n$ components $(i_1,...,i_n)$ are not necessarily ordered. We may, by
construction in the recurrence relation, demand $i_1 \geq i_2 \geq
\cdots \geq i_{n-2}$, but if we want all possible equations, we have
to let $i_{k}$ to take values bigger than $i_{k-1}$. Since this would
lead to a partition which is not an ordered one, it is necessary to
rearrange the indexes. For $\gamma$ even, this reordering does not
change the value of the coefficient (see equation
\ref{ecu:change_sign_permutation_indexes}); this is, $c_{i_1,..,i_n}$
is the same as $c_{\mu_1,...,\mu_n}$ with $(\mu_1,...,\mu_n)$ an
ordering of $(i_1,...,i_n)$. On the other hand, for $\gamma$ odd, we
get $c_{\mu_1,...,\mu_n} = (-1)^{\alpha}c_{i_1,...,i_n}$, where
$\alpha$ is the signature of the permutation needed to transform
$(i_1,...,i_n)$ into the ordered partition $(\mu_1,...,\mu_n)$.  Now,
notice that equations (\ref{ecu:recurrence_grassmann_even}) and
(\ref{ecu:recurrence_grassmann_odd}) indicate that, for $\gamma\geq
2$, any coefficient with the form
\begin{equation}
c_{i_1,...,i_{n-2},0,0}=c_{i_1,...,i_{n-2},1,0}=0
\end{equation}
since this is the only term in the expansion. From this, we deduce that the recurrence equations may be written from $K=2$. It may be useful to keep the recurrence relations for values of $x$ different than zero or one. In these cases, for $K$ even, the term $c_{i_1,...,i_{n-2},K/2,K/2}$ does not appear. This would imply that for $\gamma > 2$ and $K=2$, all the coefficients are zero. Effectively, the equations would be:
\begin{eqnarray}
&c_{i_1,...,i_{n-1},2,0}=-\frac{1}{2}c_{i_1,...,i_{n-1},1,1}\\
&4c_{i_1,...,i_{n-1},2,0}=0.
\end{eqnarray}

Where we have used $x=2$ for the second line\footnote{This is, in the case of $\gamma$ even bigger than 2. If $\gamma$ is odd, the things are much easier, since the term $c_{i_1,...,i_{n-2},K/2,K/2}$ does not even appear.}. Furthermore, if $K$ is lower than $\gamma$, we would have $\gamma/2$ (for $\gamma$ even) or $(\gamma-1)/2$ (for $\gamma$ odd) ---one for every value of $x$--- homogeneous equations to be satisfied by $K/2$ (for $K$ even) or $(K+1)/2$ (for $K$ odd) coefficients, and then, these coefficients would be zero. This is evident from all the cases, except for $\gamma$ and $K$ odd. In this case, we have $(\gamma-1)/2$ equations, and $(K+1)/2$ coefficients. However, note that since $\gamma$ is odd, the biggest odd number lower than $\gamma$ is $\gamma - 2$. In this case, we have $(\gamma-1)/2$ coefficients, and they are still all zero. So, for any sum, $K$ can start from $\gamma$.
In the same way, we may see that if $K>\gamma(2n-3)$, we would have
more homogeneous equations than coefficients ---since any component of
the partition may at most be equal to $\gamma(n-1)$--- so the maximum value of $K$ is $\gamma(2n-3)$.
We may summarize what has been said until now in this: for $\gamma$ even, we have a set of equations of the form:
\begin{equation}
 \sum_{\substack{i_{n} < i_{n-1}\\i_{n-1}+i_{n}=K}}(i_{n-1}-i_n)^xc_{i_1,...,i_{n-2},i_{n-1},i_n} = -\delta_{x,0}\frac{1}{2}c_{i_1,...,i_{n-2},K/2,K/2} \hspace{1cm} x=0,2,...,\gamma-2.
 \label{ecu:general_equation_gamma_even} 
\end{equation}

In the case of $\gamma$ odd:
\begin{equation}
 \sum_{\substack{i_{n} < i_{n-1}\\i_{n-1}+i_{n}=K}}(i_{n-1}-i_n)^xc_{i_1,...,i_{n-2},i_{n-1},i_n} = 0 \hspace{1cm} x=1,3,...,\gamma-2 
 \label{ecu:general_equation_gamma_odd}
\end{equation}
with, in both cases, $K$ running from $\gamma$ to $\gamma(2n-3)$.
 
Since for a given $x$ the equations for different values of $K$ are
highly coupled, it may be more convenient to work with all the values
of $x$ within a single value of $K$. We may start from the root
partition $(\gamma(n-1),\gamma(n-2),...,0)$ and from the
composition obtained by reordering the root partition\footnote{These two compositions belong
  to the limiting cases $K=\gamma$ and $K=\gamma(2n-3)$}
$(\gamma(n-3),\gamma(n-4),...,0,\gamma(n-1),\gamma(n-2))$. Sometimes,
abusing the language, we shall call modified root partition the
composition obtained by reordering the parts of the root partition. From both partitions, we have $\gamma/2$
equations ($(\gamma-1)/2$ if $\gamma$ is odd)  and $\gamma/2+1$
coefficients ($(\gamma+1)/2$ if $\gamma$ is odd). We lack one equation to solve the system. This can be obtained from the normalization condition (\ref{ecu:coefficients_gamma_series}) and from the  permutation sign rule (\ref{ecu:change_sign_permutation_indexes}). 

Notice here that the modified terms ---the last two terms--- in the partitions originally differ by $\gamma$. The next set of equations will correspond to the partitions in which the modified terms differ by $\gamma+1$. There will be, then, $\gamma/2+1$ coefficients ($(\gamma+1)/2+1$ if $\gamma$ is odd) in each sum. One more equation is needed in the even case, two more in the odd case. In both cases, these equations are supplied by the solutions of the first set of equations. Effectively, running from $\gamma$ to $\gamma/2$ we get $(...,2\gamma,\gamma,0) \rightarrow (...,2\gamma,\gamma-1,1)$ so the difference between the $n-2$ and the $n-1$ terms is $\gamma+1$ and we have one of the terms in the sum. In the same way, $(\gamma(n-3),...,\gamma(n-1),\gamma(n-2)) \rightarrow (\gamma(n-3),...,\gamma(n-1)-1,\gamma(n-2)+1)$, and the last and the first term differ by $\gamma+1$. Therefore, we can construct two new sets of equations from two of the coefficients calculated in the first part. Moreover, these new sets may as well define other sets, since in its construction new terms in which two components of the partition differing by $\gamma+1$ may appear. 

\subsection{Recurrence relation for the coefficients obtained from the
theory of Jack polynomials}

On the other hand, it has been observed that the products in equations
(\ref{ecu:product_Gamma_even}) and (\ref{ecu:product_Gamma_odd}) can
be expressed in terms of the Jack symmetric or antisymmetric
polynomials \cite{aana,evp}. Effectively, the powers of the Vandermonde
determinant can be written as:
\begin{eqnarray}
&\prod_{1 \leq i < j \leq n} (z_i - z_j)^{2p}=P_{2p\delta n}\left(z;-2/(2p-1)\right)\\
&\prod_{1 \leq i < j \leq n} (z_i - z_j)^{2p+1}=S_{(2p+1)\delta n}\left(z;-2/(2p+1)\right)
\end{eqnarray} 
where  $P_{\kappa}(z;\alpha)$ and $S_{\kappa}(z;\alpha)$
denote
the Jack symmetric and antisymmetric polynomials, with
$z=(z_1,\ldots,z_n)$ and $\delta n = (n-1,n-2,...0)$. 
In the case of the even power of the Vandermonde determinant, the
partition $2p\delta n=(2p(n-1),2p(n-2),\ldots,0)$ is the root
partition $\kappa$ related to the value of the coupling $\gamma=2p$
(see eq.~(\ref{ecu:root_partition})). The same
applies for the case of the odd power of the Vandermonde determinant,
$(2p+1)\delta n$ is the root partition $\kappa$ for $\gamma=2p+1$.

The symmetric and antisymmetric Jack polynomials admit an expansion of
the form
\begin{equation}
P_{\kappa}(z;\alpha)=m_{\kappa}(z)+\sum_{\rho<\kappa}c_{\kappa\rho}m_{\rho}(z)
\label{ecu:expansion_symmetric_jack}
\end{equation}
and 
\begin{equation}
S_{\kappa}(z;\alpha) = q_{\kappa}(z)+\sum_{\rho <
  \kappa} \text{\~{c}}_{\kappa\rho}q_{\rho}(z)
\,.
\label{ecu:expansion_non_symetric_jack}
\end{equation}
In the previous equations, $\rho$ are partitions obtained by successive
squeezing operations performed on $\kappa$ and $\rho < \kappa$
implying dominance.  We recall that the dominance relation between
partitions is defined by : $\rho < \kappa$ if and only if
$\sum_{i=1}^k \rho_i \leq \sum_{i=1}^k \kappa_i$ for all
$k\in\{1,\ldots, n\}$. The squeezing operation on a partition $\mu$
consists in changing $\mu_{i}\mapsto \mu_{i}- r$ and $\mu_{j} \mapsto
\mu_{j} +r$ for $i<j$.

It is known that the Jack polynomials are eigenfunctions of the operator
\cite{dfqh, forresterbook}
\begin{equation}
\mathcal{H}=\sum_i\left(z_i\frac{\partial}{\partial z_i}\right)^2+\frac{n-1}{\alpha}\sum_{i}z_i\frac{\partial}{\partial z_i}+\frac{2}{\alpha}\sum_{i<j}\frac{z_iz_j}{z_i-z_j}\left(\left(\frac{\partial}{\partial z_i}-\frac{\partial}{\partial z_j}\right)-\frac{1-M_{ij}}{z_i-z_j}\right)
\end{equation}
with $M_{ij}=1$ for the symmetric Jack polynomials and $M_{ij}=-1$ for
the antisymmetric Jack polynomials. By applying this operator on both
sides of equations (\ref{ecu:expansion_symmetric_jack}) and
(\ref{ecu:expansion_non_symetric_jack}) one can obtain a recurrence
relation for the coefficients $c_{\kappa\rho}$ and
$\text{\~c}_{\kappa\rho}$. For $\Gamma=4p$ (this is, for $\gamma=2p$
even) it is~\cite{mcdonaldbook}
\begin{equation}
c_{\kappa\rho}=\frac{1}{e_{\kappa}(\alpha)-e_{\rho}(\alpha)}\frac{2}{\alpha}\sum_{\rho < \mu \leq \kappa}\left((\rho_i+r)-(\rho_j-r)\right)c_{\kappa\mu}
\label{ecu:recurrence_relation_2_even}
\end{equation}
where $\alpha=-2/(2p-1)$ and $\mu$ is constructed from $\rho$ by
unsqueezing, i.e.~adding $r$ to its $i$-th element and subtracting $r$ to its $j$-th element. The term $e_{\kappa}(\alpha)$ is defined as:
\begin{equation}
e_{\kappa}(\alpha)=\sum_{i=1}^{n}\kappa_i\left(\kappa_i-1-\frac{2}{\alpha}(i-1)\right).
\end{equation}
For $\Gamma=4p+2$ ($\gamma=2p+1$ odd) the recurrence relation was
derived in~\cite{aana} 
\begin{equation}
\text{\~c}_{\kappa\rho}=\frac{1}{e^F_{\kappa}(\alpha)-e^F_{\rho}(\alpha)}\frac{2}{\alpha}\sum_{\rho < \mu \leq \kappa}\left(\rho_i-\rho_j\right)\text{\~c}_{\kappa\mu}(-1)^{N_{SW}}
\label{ecu:recurrence_relation_2_odd}
\end{equation}
with:
\begin{equation}
e^F_{\kappa}(\alpha)=\sum_{i=1}^{n}\kappa_i\left(\kappa_i+2i\left(1-\frac{1}{\alpha}\right)\right).
\end{equation}
Here, $(-1)^{N_{SW}}$ is the sign of the permutation needed to make $\mu$ an ordered partition, and $\alpha=-2/(2p+1)$. For the work done here, it is worth the effort to calculate explicitly the constant accompanying the sum for the case where one does an squeezing in the last two or the first two terms in the root partition; if $\gamma$ is even, if one constructs $\rho$ by squeezing the last two terms of the root partition $\kappa$ (adding $s$ to $\kappa_j$ and substracting $s$ to $\kappa_i$, with $j > i$):
\begin{equation}
e_{\kappa}(\alpha)-e_{\rho}(\alpha) = s\left(\kappa_i-s-1-\frac{2}{\alpha}\left[i-1\right]\right)-s\left(\kappa_j+s-1-\frac{2}{\alpha}\left[j-1\right]\right)+\kappa_is-\kappa_js.
\end{equation} 
Since $\kappa$ is the root partition, $\kappa_i-\kappa_j = (j-i)\gamma$. Then
\begin{equation}
e_{\kappa}(\alpha)-e_{\rho}(\alpha) = -s\left(2s+(i-j)\left[2\gamma+\frac{2}{\alpha}\right]\right).
\end{equation}
Replacing $2/\alpha = 1-\gamma$ we have, finally:
\begin{equation}
\frac{1}{e_{\kappa}(\alpha)-e_{\rho}(\alpha)}\frac{2}{\alpha} = \frac{\gamma-1}{s\left(2s+(i-j)\left[\gamma+1\right]\right)}
\label{ecu:constant_squeezed_root_even}
\end{equation}
if we have two consecutive terms, $i-j=-1$. On the other hand, for $\gamma$ odd we have, doing the same procedure:
\begin{equation}
\frac{1}{e^F_{\kappa}(\alpha)-e^F_{\rho}(\alpha)}\frac{2}{\alpha} = \frac{\gamma}{s\left(2s+(i-j)\left[\gamma-2\right]\right)}.
\label{ecu:constant_squeezed_root_odd}
\end{equation}
Here we have replaced $\alpha = -2/\gamma$. It is very likely that by
an appropriate linear transformation of
equations~(\ref{ecu:general_equation_gamma_even}) one can obtain the
relations (\ref{ecu:recurrence_relation_2_even}), and similarly in the odd case with the system of
equations (\ref{ecu:general_equation_gamma_odd}) and
(\ref{ecu:recurrence_relation_2_odd}). This conjecture is explored
with an explicit example in Sec.~\ref{sec:example}.

\subsection{The product rule}

Bernevig and Regnault have found a product rule which is fulfilled by certain partitions squeezed from the root partition \cite{aana}. Their treatment is written in terms of the occupation number of each number in the partition. In this approach the partition is written in terms of the number of times each value is repeated in it. Consider an ordered partition of the form
\begin{equation}
\mu = (\mu_1,\mu_1,...,\mu_2,\mu_2,...,\mu_n).
\end{equation}
This partition can also be written as
\begin{equation}
\mu=[\lambda_{\mu_1},...,\lambda_1,\lambda_0]
\end{equation}
where $\lambda_i$ is the number of times the number $i$ appears in the partition, and we stop when we reach the maximum value of $i$ for which $\lambda_i$ is not zero. So, for example, the root partition for $n=7$ and $\gamma=2$ can be written as
\begin{equation}
\mu=(12,10,8,6,4,2,0)=[1010101010101].
\end{equation}
We use square brackets when dealing with the occupancy
number-representation of a partition. 

Proofs of the product rule, recalled below, can be found in \cite{aana} and \cite{dfqh}, using the properties of the Jack polynomials. Here, we intend to demonstrate the relation using the expression for the expansion coefficients derived in (\ref{ecu:expression_coefficients_grassmann_integral}). The rule says as follows: ``suppose a partition that has been obtained by squeezing the root partition and that has the special property that two parts of the partition can be identified as squeezed from individual root partitions of systems with the same $\gamma$ but a smaller number of particles. Then, the coefficient corresponding to that partition is the product of the coefficients of the other two smaller partitions'' \cite{aana}. In order to prove this statement with the equations developed here, it is necessary to translate the property to the state-representation of the partition. Consider a partition $\mu$ which has been squeezed from the root partition:
\begin{equation}
\mu = (\mu_1,\mu_2,...,\mu_n).
\end{equation}
The property of ``separability'' mentioned in the occupancy
number-representation takes the following form: let the partition $\mu$ be
divided in two parts, $\mu(A)=(\mu_1,...,\mu_{m})$ and
$\mu(B)=(\mu_{m+1},...,\mu_{n})$. Now we consider the modified
partition
\begin{equation}
\mu'(A)=(\mu_1-\gamma(n-m),...,\mu_m-\gamma(n-m))
\,.  
\end{equation}
Suppose that both $\mu'(A)$ and $\mu(B)$ have been obtained from
squeezing operations of two root partitions of $m$ and $n-m$ elements
$\kappa(A)=((m-1)\gamma,(m-2)\gamma,\ldots,\gamma,0)$ and
$\kappa(B)=((n-m-1)\gamma,(n-m-2)\gamma,\ldots,\gamma,0)$. Then the coefficient $c_{\mu}$ is just
$c_{\mu}=c_{\mu'(A)}c_{\mu(B)}$. 

The reason for considering $\mu'(A)$
instead of $\mu(A)$ is evident from the occupancy
number-representation. Consider, for the sake of clarity, the
partition $[1000110001]=(9,5,4,0)$. Let us forget for a moment the
condition for its divisions (the condition being that both new
partitions have to be squeezed from a root partition themselves), and divide it in $[10001]$ and $[10001]$. It is evident, then, that these two partitions correspond to the same state-representation $(4,0)$ and not to two different partitions $(9,5)$ and $(4,0)$. This is because, when the two partitions are divided, the occupancy numbers in the left-most one correspond now to the occupancy of the \textit{original number minus the number of states represented in the right-most partition}. If the right-most partition has $m$ positions, then the occupancy number of the left most partition correspond to the occupancy of the original state minus $m$.

Now that the property has been stated in the same language as the one used in equation (\ref{ecu:expression_coefficients_grassmann_integral}), we can prove it. Consider first the two partitions $\mu'(A)$ and $\mu(B)$. The first one represents a system with the same value of $\gamma$ as the original partition $\mu$ and with $m$ particles, with $m \leq n$. The second one has the same value of $\gamma$ too, and $n-m$ particles. Therefore the coefficients for each partition are
\begin{eqnarray}
c_{\mu'(A)}=\int _{\Lambda_1}\prod_{k=1}^{m}d\psi_k^{(1)}\cdots d\psi_k^{(\gamma)}\Xi(\mu'_k(A)+\gamma;\psi) \\
c_{\mu(B)}=\int _{\Lambda_2}\prod_{k=1}^{n-m}d\xi_k^{(1)}\cdots d\xi_k^{(\gamma)}\Xi(\mu_k(B)+\gamma;\xi).
\end{eqnarray}
Calculating their product
\begin{equation}
c_{\mu'(A)}c_{\mu(B)}=\int_{\Lambda}\left(\prod_{k=1}^{m}d\psi_k^{(1)}\cdots d\psi_k^{(\gamma)}\Xi(\mu'_k(A)+\gamma;\psi)\right)\left(\prod_{\ell=1}^{n-m}d\xi_{\ell}^{(1)}\cdots d\xi_{\ell}^{(\gamma)}\Xi(\mu_{\ell}(B)+\gamma;\xi)\right).
\end{equation}
The important portion here is the term $\Xi(J^{(\ell)};\xi)\Xi(Q^{(k)};\psi)$ with $J^{(\ell)}=\mu_{\ell}(B)+\gamma$ and $Q^{(k)}=\mu'_{k}(A)+\gamma$. Expanding it with the representation in (\ref{ecu:expression_coefficients_sum_delta}) of the function $\Xi$ we have
\begin{equation}
\Xi(J^{(\ell)};\xi)\Xi(Q^{(k)};\psi) = \sum_{j_{1}^{(\ell)},...,j_{\gamma}^{(\ell)}=1}^{n-m}\sum_{q_{1}^{(k)},...,q_{\gamma}^{(k)}=1}^{m}\xi_{j_{1}^{(\ell)}}^{(1)}\cdots\xi_{j_{\gamma}^{(\ell)}}^{(\gamma)}\psi_{q_{1}^{(k)}}^{(1)}\cdots\psi_{q_{\gamma}^{(k)}}^{(\gamma)}\delta_{j_{1}^{(\ell)}+\cdots+j_{\gamma}^{(\ell)},J^{(\ell)}}\delta_{q_{1}^{(k)}+\cdots+q_{\gamma}^{(k)},Q^{(k)}}.
\end{equation}
Let us do a simple change of variable: let each $q_i^{(k)}=p_i^{(k)}-(n-m)$. Then, let each $\psi^{(i)}_{q_{i}^{(k)}}=\xi^{(i)}_{q_{i}^{(k)}+n-m}=\xi^{(i)}_{p_{i}^{(k)}}$. The last expression becomes
\begin{eqnarray}
&&\Xi(J^{(\ell)};\xi_1)\Xi(Q^{(k)};\xi_2) = \notag \\
&& \hspace{1.5cm} \sum_{j_{1}^{(\ell)},...,j_{\gamma}^{(\ell)}=1}^{n-m}\sum_{p_{1}^{(k)},...,p_{\gamma}^{(k)}=n-m+1}^{n}\xi_{j_{1}^{(\ell)}}^{(1)}\cdots\xi_{j_{\gamma}^{(\ell)}}^{(\gamma)}\xi_{p_{1}^{(k)}}^{(1)}\cdots\xi_{p_{\gamma}^{(k)}}^{(\gamma)}\delta_{j_{1}^{(\ell)}+\cdots+j_{\gamma}^{(\ell)},J^{(\ell)}}\delta_{p_{1}^{(k)}+\cdots+p_{\gamma}^{(k)},Q^{(k)}+\gamma(n-m)}. \notag
\end{eqnarray}
Now, notice that $\mu(B)$ is a partition of the number $\gamma m (m-1)/2$ and then $\gamma \leq J^{(\ell)} \leq \gamma m$. On the other hand, $\mu'(A)$ is a partition of the number $\gamma (n-m)(n-m-1)/2$, and then, $\gamma \leq Q^{(k)} \leq \gamma (n-m)$. Then the limits of both sums can be rewritten ---since the $\delta$ terms will get rid of most of the factors---, and
\begin{eqnarray}
&&\Xi(J^{(\ell)};\xi)\Xi(P^{(k)};\xi) = \notag \\
&& \hspace{1.5cm} \sum_{j_{1}^{(\ell)},...,j_{\gamma}^{(\ell)}=1}^{n}\xi_{j_{1}^{(\ell)}}^{(1)}\cdots\xi_{j_{\gamma}^{(\ell)}}^{(\gamma)}\delta_{j_{1}^{(\ell)}+\cdots+j_{\gamma}^{(\ell)},J^{(\ell)}}\sum_{p_{1}^{(k)},...,p_{\gamma}^{(k)}=1}^{n}\xi_{p_{1}^{(k)}}^{(1)}\cdots\xi_{p_{\gamma}^{(k)}}^{(\gamma)}\delta_{p_{1}^{(k)}+\cdots+p_{\gamma}^{(k)},P^{(k)}}
\end{eqnarray}
where we have defined a new variable $P^{(k)}=Q^{(k)}+\gamma(m-n)$. Finally, then, we have
\begin{equation}
c_{\mu'(A)}c_{\mu(B)}=\int_{\Lambda}\left(\prod_{k=n-m}^{n}d\xi_k^{(1)}\cdots d\xi_k^{(\gamma)}\Xi(P^{(k)};\xi)\right)\left(\prod_{\ell=1}^{n-m}d\xi_{\ell}^{(1)}\cdots d\xi_{\ell}^{(\gamma)}\Xi(J^{(\ell)};\xi)\right).
\end{equation}
Coming back to the partitions $\mu'(A)$ and $\mu(B)$ and reorganizing, we see that $P^{(k)}=\mu'_k(A)+\gamma(n-m)+\gamma=\mu_k(A)+\gamma$ and
\begin{equation}
c_{\mu'(A)}c_{\mu(B)}=\int_{\Lambda}\prod_{k=1}^{n}d\xi_{k}^{(1)}\cdots d\xi_{k}^{(\gamma)}\Xi(\mu_k+\gamma;\xi) = c_{\mu}
\end{equation}
which completes the proof. 

\section{An Instructive Example}
\label{sec:example}

As an example of the previous analysis, we can calculate the coefficients for the case $\gamma=4$, $n=3$. Here, $K$ runs from $4$ to $12$. The maximum value a term in the partition can take is $8$ and the number to be partitioned is $12$. The root partition is $(8,4,0)$ and the first set of equations~(\ref{ecu:general_equation_gamma_even}) are:
\begin{eqnarray*}
c_{8,4,0}+c_{8,3,1}=-c_{8,2,2}/2\\
16c_{8,4,0}+4c_{8,3,1}=0\\
c_{8,4,0}=1
\end{eqnarray*}    
so $c_{8,3,1}=-4$ and $c_{8,2,2}=6$. And:
\begin{eqnarray*}
c_{0,8,4}+c_{0,7,5}=-c_{0,6,6}/2\\
16c_{0,8,4}+4c_{0,7,5}=0\\
c_{0,8,4}=1
\end{eqnarray*}
so $c_{0,7,5} = c_{7,5,0}=-4$ and $c_{0,6,6} = c_{6,6,0} = 6$. We now pick from these equations the coefficients with terms which differ in $\gamma+1=5$, and construct a new set of equations:
\begin{eqnarray*}
c_{8,3,1}+c_{7,4,1}+c_{6,5,1} =0\\
25c_{8,3,1}+9c_{7,4,1}+c_{6,5,1} =0\\
c_{8,3,1} = -4
\end{eqnarray*}
so $c_{7,4,1}=12$ and $c_{6,5,1}=-8$. Also:
\begin{eqnarray*}
c_{7,5,0}+c_{7,4,1}+c_{7,3,2}=0\\
25c_{7,5,0}+9c_{7,4,1}+c_{7,3,2}=0\\
c_{7,5,0}= -4
\end{eqnarray*}
which throws $c_{7,3,2}=-8$. However, here we note that there are two more partitions with terms that differ by $5$: $(6,5,1)$ and $(7,3,2)$. For these partitions too we construct the set of equations:
\begin{eqnarray*}
c_{7,5,0}+c_{6,5,1}+c_{5,5,2}+c_{4,5,3} =0\\
49c_{7,5,0}+25c_{6,5,1}+9c_{5,5,2}+c_{4,5,3} =0\\
c_{6,5,1} = -8\\
c_{7,5,0} = -4
\end{eqnarray*}
and then $c_{5,4,3}=-36$ and $c_{5,5,2}=48$. On the other hand:
\begin{eqnarray*}
c_{8,3,1}+c_{7,3,2}+c_{6,3,3}+c_{5,3,4} =0\\
49c_{8,3,1}+25c_{7,3,2}+9c_{6,3,3}+c_{5,3,4} =0\\
c_{7,3,2} = -8\\
c_{8,3,1} = -4
\end{eqnarray*}
so $c_{6,3,3}=48$. Now, from all the coefficients found before, we pick the ones with terms which differ by $\gamma+2=6$. These are those of the partitions $(8,2,2)$, $(6,6,0)$ and $(7,4,1)$. The set of equations for them are:
\begin{eqnarray*}
c_{8,2,2}+c_{7,3,2}+c_{6,4,2}=-c_{5,5,2}/2\\
36c_{8,2,2}+16c_{7,3,2}+4c_{6,4,2}=0\\
c_{8,2,2} = 6\\
c_{7,3,2} = -8
\end{eqnarray*}
then $c_{6,4,2}=-22$ and $c_{5,5,2}=48$. For the second partition we get:
\begin{eqnarray*}
c_{6,6,0}+c_{6,5,1}+c_{6,4,2}=-c_{6,3,3}/2\\
36c_{6,6,0}+16c_{6,5,1}+4c_{6,4,2}=0\\
c_{6,6,0} = 6\\
c_{6,5,1} = -8
\end{eqnarray*}
which has only known terms (the reader can check that the values derived from this equations are the same already found). Finally, for the third partition:
\begin{eqnarray*}
c_{8,4,0}+c_{7,4,1}+c_{6,4,2}+c_{5,4,3}=-c_{4,4,4}/2\\
64c_{8,4,0}+36c_{7,4,1}+16c_{6,4,2}+4c_{5,4,3}=0\\
c_{7,4,1} = 12\\
c_{5,4,3} = -36\\
c_{8,4,0} = 1
\end{eqnarray*}
and we get $c_{4,4,4}=90$. The results are summarized in table \ref{table:coefficients_g=4_n=3}. 

\begin{table}[htb]
\centering
\begin{tabular}{|c|c|}
\hline
Partition & Coefficient ($c_{\mu}$)\\
\hline
$[8,4,0]$ & 1\\
$[8,3,1]$ & -4\\
$[8,2,2]$ & 6\\
$[7,5,0]$ & -4\\
$[7,4,1]$ & 12\\
$[7,3,2]$ & -8\\
$[6,6,0]$ & 6\\
$[6,5,1]$ & -8\\
$[6,4,2]$ & -22\\
$[6,3,3]$ & 48\\
$[5,5,2]$ & 48\\
$[5,4,3]$ & -36\\
$[4,4,4]$ & 90\\
\hline
\end{tabular}\caption{Expansion Coefficients for the Partition Function of the 2dOCP, with $\Gamma=8$ and $n=3$}
\label{table:coefficients_g=4_n=3}
\end{table}

It is worth to note that there is a symmetry for the coefficients of certain sets. The mid value for $K$ is $\gamma(n-1)$. The equations obtained for $K=\gamma(n-1)+\rho$ and $K=\gamma(n-1)-\rho$ are the same. This happens because the original equations for the root partition and the modified root partition ---which have $K=\gamma$ and $K=\gamma(2n-3)$, and hence are at the same distance from $\gamma(n-1)$--- are always the same. Effectively, the first two sets of equations can be written in matrix form as:
\begin{equation}
\left(\begin{array}{ccccc}
1 & 1 & \cdots & 1 & 1/2\\
(\gamma-2)^2 & (\gamma-4)^2 & \cdots & 4 & 0\\
\vdots & \vdots & \ddots & \vdots & \vdots\\
(\gamma-2)^{\gamma-2} & (\gamma-4)^{\gamma-2} & \cdots & 2^{\gamma-2} & 0
\end{array}\right)\left(\begin{array}{c}
c_{...,i\gamma-1,j\gamma+1...}\\
c_{...,i\gamma-2,j\gamma+2,...}\\
\vdots\\
c_{...,K/2,K/2,...}
\end{array}\right)=\left(\begin{array}{c}
-1\\
-\gamma^2\\
\vdots\\
-\gamma^{\gamma-2}
\end{array}\right)
\label{ecu:matrices_first_coeff_gamma_even}
\end{equation}  

Where $K=\gamma$ or $K=\gamma(2n-3)$, $i=1$ and $j=0$, or $i=(n-1)$ and $j=(n-2)$ and for $\gamma$ even. For $\gamma$ odd:
\begin{equation}
\left(\begin{array}{cccc}
(\gamma-2) & (\gamma-4) & \cdots & 1 \\
(\gamma-2)^3 & (\gamma-4)^3 & \cdots & 1\\
\vdots & \vdots & \ddots & \vdots\\
(\gamma-2)^{\gamma-2} & (\gamma-4)^{\gamma-2} & \cdots & 1
\end{array}\right)\left(\begin{array}{c}
c_{...,i\gamma-1,j\gamma+1...}\\
c_{...,i\gamma-2,j\gamma+2,...}\\
\vdots\\
c_{...,(K+1)/2,(K-1)/2,...}
\end{array}\right)=\left(\begin{array}{c}
-\gamma\\
-\gamma^3\\
\vdots\\
-\gamma^{\gamma-2}
\end{array}\right)
\label{ecu:matrices_first_coefficients_gamma_odd}
\end{equation}

Hence, the next set of equations will again be the same, since we will have a collection of coefficients which differ by the same amount: in the $\gamma$ even case, we will run over the components $\gamma(n-2)+1,\gamma(n-3)$ and $3\gamma,2\gamma-1$. The differences among them are the same ($\gamma+1$), so their coefficients will have the same constant in every equation; further, they differ in the same amount, so both equations will have the same number of terms. This will hold for further sets, so it suffices to calculate the coefficients for $K$ up to $\gamma(n-1)$, and then to equate the missing ones for $K > \gamma(n-1)$ to their counterparts. The matrices for the sets of equations for values of $K$ bigger than $\gamma$ will have the same form of those in (\ref{ecu:matrices_first_coeff_gamma_even}) and (\ref{ecu:matrices_first_coefficients_gamma_odd}), provided the adequate coefficients are put in the right side of the expression. 

Looking for the connection between the two recurrence relations mentioned before may be hard for an arbitrary partition. However, for partitions squeezed from the root partition, the constant terms in (\ref{ecu:recurrence_relation_2_even}) and (\ref{ecu:recurrence_relation_2_odd}) are quite simple (see equations \ref{ecu:constant_squeezed_root_even} and \ref{ecu:constant_squeezed_root_odd}). If we use them for our example, we get:
\begin{equation}
c_{8,3,1}=-1(4-0)c_{8,4,0}=-4c_{8,4,0} = -4
\end{equation}

Which is precisely the first equation derived. Now:
\begin{equation}
c_{8,2,2}=-\frac{3}{2}\left(4c_{8,4,0}+2c_{8,3,1}\right) = 6
\end{equation}

This second equation is a lineal combination of the first two:
\begin{equation}
-6c_{8,4,0}-3c_{8,3,1} = -\frac{1}{4}\left(16c_{8,4,0}+4c_{8,3,1}\right)-2\left(c_{8,4,0}+c_{8,3,1}\right)
\end{equation}

So, the recurrence relation presented in \cite{evp, aana,dfqh} can be expressed
as a linear combination of the equations derived by \v{S}amaj~\cite{i2ds}. More generally, the set of equations for the coefficients generated by squeezing the fisrt two or the last two terms in the root partition will look like (for $\gamma$ even):
\begin{equation}
\left(\begin{matrix}
1 & 0 & \cdots & 0\\
-\tau_2 (\gamma-2) & 1  & \cdots & 0\\
\vdots & \vdots & \ddots & \vdots \\
-\tau_{\gamma/2} (\gamma-2) & -\tau_{\gamma/2} (\gamma-4) & \cdots & 1  
\end{matrix}\right)\left(\begin{matrix}
c_{...,i\gamma-1,j\gamma+1,...}\\
c_{...,i\gamma-2,j\gamma+2,...}\\
\vdots\\
c_{...,K/2,K/2,...}
\end{matrix}\right)=\left(\begin{matrix}
\tau_1\\
\tau_2\\
\vdots\\
\tau_{\gamma/2}
\end{matrix}\right)\gamma
\end{equation}
with, again, $K=\gamma$ or $K=(2n-3)\gamma$. Here, $\tau_{s}$ is:
\begin{equation}
\tau_{s} = \frac{\gamma-1}{s\left(2s-\left[\gamma+1\right]\right)}.
\end{equation}
This is so because the only partitions which dominate $\left(\kappa_1-s,\kappa_2+s,...,\kappa_n\right)$ and $\left(\kappa_1,\kappa_2,...,\kappa_{n-1}-s,\kappa_n+s\right)$ are partitions of the same kind, $\left(\kappa_1-s',\kappa_2+s',...,\kappa_n\right)$ and $\left(\kappa_1,\kappa_2,...,\kappa_{n-1}-s',\kappa_n+s'\right)$ with $s'<s$. For $\gamma$ odd we have
\begin{equation}
\left(\begin{matrix}
1 & 0 & \cdots & 0\\
-\tau^F_2(\gamma-2) & 1  & \cdots & 0\\
\vdots & \vdots & \ddots & \vdots \\
-\tau^F_{(\gamma-1)/2} & -\tau^F_{(\gamma-1)/2} & \cdots & 1  
\end{matrix}\right)\left(\begin{matrix}
c_{...,i\gamma-1,j\gamma+1,...}\\
c_{...,i\gamma-2,j\gamma+1,...}\\
\vdots\\
c_{...,(K+1)/2,(K-1)/2,...}
\end{matrix}\right)=\left(\begin{matrix}
\tau^F_1\gamma\\
\tau^F_2(\gamma-2)\\
\vdots\\
\tau^F_{(K+1)/2}
\end{matrix}\right)
\end{equation}
with:
\begin{equation}
\tau^F_{s}=\frac{\gamma}{s\left(2s-\left[\gamma-2\right]\right)}
\end{equation}

We see that the relation between the two methods is not trivial, as it
is difficult to find the explicit linear transformation between one
system of equations and another, for arbitrary values of $\gamma$ and
$n$. However, for the example considered in this section, the
existence of a relation was explicitly shown. However, the question is
still open if this relation can be stated explicitly for any group of
coefficients.

\section{Concluding Remarks}

The main result of the present work consisted on showing the equivalence between the two representation for the partition function of the 2dOCP, for even values of the coupling constant $\Gamma = q^2/(2\pi\varepsilon k_BT)$. The expansion of the Vandermonde determinant as a product of Grassman variables' integrals led to another expression for the coefficients $c_{\mu}$ in the expansion (\ref{ecu:partition_meth_1_general}). In doing so, another set of equations involving these coefficients were shown to appear, as first derived by \v{S}amaj in \cite{i2ds}. A parity property was demonstrated, as well as the know product rule for the coefficients, first derived by Bernevig and Regnault in \cite{aana}. An analysis on some specific partitions led to the result that the complete set of equations was smaller than first thought, since the value of the constant $K$ in equations (\ref{ecu:recurrence_grassmann_even}) and (\ref{ecu:recurrence_grassmann_odd}) does not have to run over all values from $0$ to $\gamma(n-1)$.

A simple case was worked out, where all the properties derived could
be appreciated. Here, it became notorious that the sets of equations
derived from the Grassman representation are related to the set of
equations presented in \cite{aana} and briefly mentioned in the last
part of section 4.1. The general expression for the first sets of
equations of both relations were presented at the end of section 5,
but it remains as an open problem the general link between the two for
arbitrary values of $\gamma$ and $n$. 

{\small \textbf{Acknowledgment\ } G.T. acknowledges partial financial
  support from ECOS Nord/COLCIENCIAS-MEN-ICETEX.}

\bibliographystyle{unsrt}
\bibliography{biblio}

\end{document}